# Jet cone size dependence of single inclusive jet suppression due to jet quenching in Pb + Pb collisions at $\sqrt{s_{NN}} = 5.02$ TeV

Qing-Fei Han,[1] Man Xie,[1,2,*] and Han-Zhong Zhang[1,†]

[1]*Key Laboratory of Quark and Lepton Physics (MOE) and Institute of Particle Physics, Central China Normal University, Wuhan 430079, China*

[2]*School of Physics and Mechanics, Wuhan University of Science and Technology, Wuhan, Hubei 430065, China*

Jet suppression in high-energy heavy-ion collisions results from jet energy loss and transverse-momentum broadening during jet propagation through the quark-gluon plasma (QGP). The jet cone size ($R$) dependence of this suppression offers crucial insights into the energy loss mechanisms and QGP transport properties. In our study, we implement a comprehensive approach within the perturbative QCD parton model that incorporates both elastic and inelastic energy loss mechanisms. For elastic processes the contribution from recoiling thermal partons reduces the net in-cone energy loss for a given jet radius. For inelastic processes, we account for the angular distribution of radiated gluons, the thermalization of soft gluons, and transverse-momentum broadening. Using this framework, we calculate the jet nuclear modification factors ($R_{AA}$) and their double ratios $R_{AA}(R = 0.2–1.0)/R_{AA}(R = 0.2)$, and systematically compare with ALICE, ATLAS, and CMS data in 0–10% and 30–50% Pb + Pb collisions at $\sqrt{s_{NN}} = 5.02$ TeV. Numerical results show that $R_{AA}$ increases with the cone size $R$ because the in-cone energy loss decreases at larger radii. Specifically, as the radius $R$ grows, the probability for elastically scattered partons to escape the jet cone and the likelihood for radiated gluons to fall outside the cone both decrease, resulting in a net reduction of energy loss. The $R_{AA}$ double ratios are approximately unity for small radii ($R = 0.4$ relative to $R = 0.2$) and at high $p_T \gtrsim 200$ GeV/$c$, in agreement with the data within uncertainties.

## I. INTRODUCTION

Relativistic heavy-ion collision experiments conducted at the Relativistic Heavy Ion Collider (RHIC) [1–7] and the Large Hadron Collider (LHC) [8–13] provide a unique environment to create a new state of matter, the quark-gluon plasma (QGP), characterized by extremely high temperature and density. In the early stage of heavy-ion collisions, rare high transverse momentum ($p_T$) partons are produced via hard scatterings. As these energetic partons traverse the hot and dense QCD medium, they undergo elastic scatterings with thermal partons and inelastic scatterings accompanied by gluon radiation, resulting in significant energy loss and transverse-momentum broadening [14–19]. After losing energy, these partons are reconstructed as jets or fragmented into hadrons, whose yields are significantly suppressed relative to those in proton-proton ($p + p$) collisions [20–30]. This phenomenon, known as jet quenching [31–33], serves as a critical hard probe of the transport properties of the QGP.

The suppression of high-$p_T$ hadron yields has been well described by various theoretical models incorporating different energy-loss formalisms [34–40], and the high precision of the experimental data has enabled quantitative constraints on jet energy-loss parameters [41–45]. However, hadron yield suppression is primarily sensitive to the average energy loss of the leading parton and thus is less discriminating among different energy-loss formalisms [29,46,47]. In contrast, inclusive jets, collimated clusters of particles reconstructed within a given cone size $R$, are sensitive not only to the leading parton's energy loss, but also to the redistribution of the lost energy within the medium via radiated gluons and recoil partons that may fall inside the jet cone. Additionally, jet observables are less sensitive to nonperturbative hadronization modeling than hadron observables [29,46,48]. As the jet cone size $R$ varies, the reconstructed jet captures different fractions of the quenched jet energy and the medium response. Consequently, the $R$ dependence of inclusive-jet suppression has attracted wide attention, as it is sensitive to QGP properties and the underlying mechanisms of jet quenching.

With increasing colliding energy, larger data statistics, and improved background-subtraction techniques, measurements for inclusive jets have become more precise and cover a wider kinematic range, enabling detailed studies of in-jet energy-loss mechanisms. The ATLAS experimental collaboration first reported measurements of single inclusive jet suppression across the jet radii from $R = 0.3$ to $R = 0.5$ in Pb + Pb collisions at $\sqrt{s_{NN}} = 2.76$ TeV [25,49]. The strength of jet suppression is quantified by the nuclear modification factor $R_{AA}$, defined as the ratio of the jet production spectrum in nucleus-nucleus ($A + A$) collisions to that in $p + p$ collisions. ATLAS also reported the double ratios of $R_{AA}$ relative to $R = 0.2$, which exhibit a weak increase with increasing $R$. These double ratios are key observables for quantifying the $R$ dependence of jet quenching. The CMS [29] and ALICE [27,50] Collaborations also measured the suppression of inclusive jet production in 2.76 TeV Pb + Pb collisions and pointed out that the jet suppression does not depend on $R$. For lower-$p_{\mathrm{T}}$ jets, differences in background-subtraction methods and jet-selection criteria may affect the jet yield measurements [29]. At $\sqrt{s_{NN}} = 5.02$ TeV, measurements of inclusive jet production cover a wider kinematic range, from about 40 GeV to 1 TeV, with higher precision [51,52]. Meanwhile, the jet radius parameter has been extended up to $R = 1.0$ [28,30]. The CMS and ALICE Collaborations continue to observe only a weak $R$ dependence of jet $R_{AA}$, whereas the ATLAS Collaboration suggests that large-$R$ jets may contain more complex substructure, leading to a stronger suppression [26]. In any case, these high-precision data provide a stringent basis for further constraining and refining jet-quenching models.

*Contact author: manxie@wust.edu.cn
†Contact author: zhanghz@mail.ccnu.edu.cn

Many recent theoretical studies of jets have investigated the jet-radius dependence of inclusive jet suppression by accounting for multiple sources of energy loss, including elastic and inelastic processes, medium response, color coherence, and initial-state cold nuclear matter effects. These mechanisms lead to distinct dependences on the jet radius. Within the soft-collinear effective theory (SCET) framework [53,54], including both cold nuclear matter and radiative energy loss yields a suppression for jets that weakens slightly with increasing $R$. The JEWEL generator [55,56], when including collisional and radiative energy loss, tends to overestimate the suppression; after incorporating medium response that restores energy within the jet, it predicts $R$-independent jet suppression. The linear Boltzmann transport (LBT) model [46,57] also takes the jet-induced medium response into account. Somewhat differently, LBT gives an $R$-independent result when only energy loss is included, while adding medium response leads to weaker suppression with increasing $R$. Both JEWEL and LBT models indicate that jet-induced medium response reduces the net energy loss of jets within a given cone size. The Hybrid Model [58], which employs a strongly coupled holographic approach based on $N = 4$ supersymmetric Yang-Mills theory to model the interaction of shower partons with the QGP, also yields $R$-independent suppression when medium response is included. Moreover, a first-principles calculation [59–61] that incorporates the latest next-to-leading order (NLO) gluon spectrum within the improved opacity expansion and color-coherence effects, analyzes the effects of resummation of energy loss effects arising from hard, vacuumlike emissions within the medium, along with modeling of soft energy flow and its recovery at the jet cone on the $R$-dependent jet suppression. This calculation indicates that fluctuations of the jet substructure can lead to stronger suppression for larger-$R$ jets, consistent with ATLAS observations [26].

In our previous study [62], within a next-to-leading order (NLO) perturbative QCD (pQCD) parton model, we investigated the effects of collisional energy loss with medium recoil partons and radiative energy loss with the $p_{\mathrm{T}}$ broadening on inclusive jet production in Pb + Pb collisions at $\sqrt{s_{NN}} = 2.76$ TeV. Collisional energy loss was calculated from elastic scattering rates, and recoil thermal partons that fall inside the jet cone do not contribute to the net jet energy loss [57]. The radiative energy loss was obtained by the higher-twist formalism [17,18,62,63], wherein only radiated gluons that exit the jet cone contribute to net energy loss. We further accounted for the transverse-momentum broadening effect, which enhances the probability of radiated gluons exiting the jet cone. This pQCD framework provided good descriptions of the jet data in Pb + Pb collisions at $\sqrt{s_{NN}} = 2.76$ TeV. Here, we integrate these mechanisms within a global analysis of inclusive jet suppression at $\sqrt{s_{NN}} = 5.02$ TeV. All available inclusive jet data within a given centrality bin are used to determine the sole unknown parameter, the strong coupling $\alpha_s$, which controls the magnitude of both collisional and radiative energy loss. With this calibrated framework, we provide a systematic description of the transverse-momentum and centrality dependence of inclusive jet suppression, with particular attention to the resulting jet-radius dependence when all above effects are included.

It should be noted that we employ a relatively naive pQCD model in which final-state partons are clustered into jets using the anti-$k_{\mathrm{T}}$ algorithm [64]. However, logarithmic terms arising from soft or collinear radiation require specialized resummation techniques, such as threshold resummation [65,66] or dedicated treatments within infrared-safe jet algorithms [67], to ensure perturbative convergence and to better describe the dependence of jet cross sections on the radius $R$ and transverse momentum $p_{\mathrm{T}}$. Recently developed semi-inclusive jet functions $J_c(z, \omega R, \mu)$ in the SCET framework [54,68], along with the inclusive microjet fragmentation function $f_{j/i}^{\mathrm{incl}}(z, t)$ derived from first principles [69], which include the resummation of $\ln R$ terms in the small-$R$ limit, have shown improved agreement with small-$R$ jet data in $pp$ collisions. In addition, nonperturbative effects such as hadronization are crucial for a more comprehensive interpretation of the $R$ dependence of jet observables [58,70,71]. In this work, we focus primarily on the net energy loss within a given $R$ arising from collisional and radiative processes and investigate whether $p_{\mathrm{T}}$ broadening tends to push radiated gluons outside the jet cone. We also examine the quantitative impact of medium response on the net energy loss. Accordingly, we adopt a simplified parton-level jet model. For consistency with the $pp$ baseline, we primarily vary the factorization scale $\mu$ to effectively account for nonperturbative contributions. We acknowledge that, in principle, the same factorization scale should be used across different jet-$R$ values. However, to better isolate medium effects, we adjust the factorization scale so that the $pp$ baselines for different jet-cone sizes better fit the experimental data. Furthermore, we subtract an average energy loss from the reconstructed jet energy, which may influence the $p_{\mathrm{T}}$ dependence of jet data in $A+A$ collisions. Nonetheless, this simplification is not expected to alter the $R$ dependence of different energy-loss mechanisms. We recognize that future studies should incorporate both jet fragmentation and hadronization to more accurately investigate the $R$ dependence of jet suppression.

The paper is organized as follows. In Sec. II, we introduce the NLO pQCD parton model and present the differential cross sections for single inclusive jets. In Sec. III, we describe the implementation of collisional energy loss with recoil thermal partons and radiative energy loss with $p_T$ broadening for jets propagating through the QGP. In Sec. IV, we present numerical results for differential cross sections, nuclear modification factors ($R_{AA}$), and double ratios of $R_{AA}$ for inclusive jets as functions of transverse momentum and centrality in Pb + Pb collisions at $\sqrt{s_{NN}} = 5.02$ TeV, and compare them with the available experimental data. A brief summary and discussion are given in Sec. V.

## II. NLO PQCD PARTON MODEL

Within the pQCD parton model, the differential cross section of single inclusive jet at leading order (LO) in high energy proton-proton ($p+p$) collisions can be obtained as follows [72]:

$$\frac{d\sigma_{pp\to \mathrm{jet}+X}}{dy\,dp_T} = 2p_T\sum_{abcd}\int dy_d x_a f_{a/p}(x_a,\mu^2) \times x_b f_{b/p}(x_b,\mu^2)\frac{d\hat{\sigma}_{ab\to cd}}{d\hat{t}}, \tag{1}$$

where $x_a = \frac{2p_T}{\sqrt{s}}(e^{y_c}+e^{y_d})$ and $x_b = \frac{2p_T}{\sqrt{s}}(e^{-y_c}+e^{-y_d})$ are the moment fractions of partons $a$ and $b$ carried from the protons, $y = y_c$ and $y_d$ are the final partons' rapidities in the $2\to 2$ scattering processes; $f_{a/p}(x_a,\mu^2)$ is the parton distribution function (PDF) for a parton $a$ with momentum fraction $x_a$ from a free nucleon, taken from CT18 parametrizations [73]. $d\hat{\sigma}_{ab\to cd}/d\hat{t}$ is the differential cross section for parton-parton hard scattering process at leading order ($\alpha_s^2$), $\hat{t} = -p_T^2(1+e^{y_d-y_c})$, along with $\hat{s} = x_a x_b s_{NN}$ and $\hat{u} = -p_T^2(1+e^{y_c-y_d})$, are the Mandelstam variables. In this study, we calculate the single inclusive jet cross sections to next-to-leading order (NLO). In the NLO correction term $\mathcal{O}(\alpha_s^3)$, we consider both $2\to 3$ real tree diagram contributions and $2\to 2$ one-loop virtual diagram contributions. In our numerical calculations, we use two cutoffs to handle the collinear singularities and soft singularities. The ultraviolet divergences can be solved by renormalization. For more detailed discussions on the NLO calculations, one can find them in Refs. [65,74].

In high-energy nucleus-nucleus ($A+A$) collisions, the single inclusive jet cross section can be written as [62],

$$\frac{dN_{AA\to \mathrm{jet}+X}}{dy\,dp_T} = \sum_{abcd}\int d^2r\, t_A(\vec{r})t_B(\vec{r}+\vec{b})\int dy_d \times\left[2p_T x_a f_{a/A}(x_a,\mu^2,\vec{r})x_b f_{b/B}(x_b,\mu^2,\vec{r}+\vec{b}) \times \frac{d\hat{\sigma}_{ab\to cd}}{d\hat{t}}\right]_{p_T\to p_T+\Delta E_{\mathrm{jet}}^{\mathrm{tot}}}, \tag{2}$$

where $t_A(\vec{r})$ is the nuclear thickness function given by a Woods-Saxon distribution [75] for nucleons in a nucleus, and is normalized as $\int d^2r t_A(\vec{r}) = A$. $f_{a/A}(x_a,\mu^2,\vec{r})$ are the nuclear modified parton distribution functions (nPDFs), which can be factorized into the parton distribution functions inside a free nucleon, $f_{a/N}(x_a,\mu^2)$, and the nuclear shadowing factor $S_{a/A}(x_a,\mu^2,\vec{r})$ [76,77],

$$f_{a/A}(x_a,\mu^2,\vec{r}) = S_{a/A}(x_a,\mu^2,\vec{r})\left[\frac{Z}{A}f_{a/p}(x_a,\mu^2) + \left(1-\frac{Z}{A}\right)f_{a/n}(x_a,\mu^2)\right], \tag{3}$$

$Z$ and $A$ are the proton number and mass number of the nucleus, respectively. The spatial dependent nuclear shadowing factor $S_{a/A}(x_a,\mu^2,\vec{r})$ can be obtained in the following form [78,79]:

$$S_{a/A}(x_a,\mu^2,\vec{r}) = 1 + [S_{a/A}(x_a,\mu^2)-1]\frac{At_A(\vec{r})}{\int d^2\vec{r}[t_A(\vec{r})]^2}, \tag{4}$$

where $S_{a/A}(x_a,\mu^2)$ is given by the EPPS21 parametrizations [80]. The CT18 PDFs and EPPS21 nPDFs are all at NLO to match the parton hard scattering cross sections.

Indeed, the cross sections in Eqs. (1) and (2) are computed at the parton level. Jets are reconstructed using the anti-$k_T$ algorithm. At LO, each back-to-back final-state parton directly forms a jet. At NLO, the anti-$k_T$ algorithm clusters one or two final-state partons into a jet. The $R$ dependence of the jet cross section in $pp$ collisions originates from the jet-clustering procedure, whereas in $A+A$ collisions it is primarily governed by the $R$ dependence of the net energy loss, e.g., whether medium-induced radiated gluons escape the jet cone (out-of-cone radiation) and related effects. We should account for jet fragmentation and its impact on the cone-size and jet-momentum dependence of energy loss in future work.

## III. ENERGY LOSS IN $A+A$ COLLISIONS

In this study we consider both the elastic and inelastic jet energy losses. In the elastic process, we take into account the influence of those soft particles that fall within the jet cone on the net jet energy loss. For the inelastic energy loss, based on higher-twist radiative energy loss, we further include the contribution of the transverse momentum ($p_T$) broadening effect.

### A. Collisional energy loss with recoil thermal partons

The collisional energy loss rate of a jet with jet cone size $R$ can be calculated from the elastic scattering rate [57,62]:

$$\frac{dE_a^{\mathrm{el}}}{dx} = \sum_{bcd}\int d\theta_2 d\theta_3 d\phi_3 dE_3 f_b(E_2,T)\frac{|\mathcal{M}_{ab\to cd}|^2}{16E_1(2\pi)^4} \times S_2(s,t,u)\frac{E_2E_3\sin\theta_2\sin\theta_3\delta E_{ab\to cd}}{E_1(1-\cos\theta_2)-E_3(1-\cos\theta_3)}, \tag{5}$$

where

$$E_2 = \frac{E_1E_3(1-\cos\theta_3)}{E_1(1-\cos\theta_{12})-E_3(1-\cos\theta_{23})}. \tag{6}$$

TABLE I. The matrix elements $|\mathcal{M}_{ij\to kl}|^2$ for two-body parton-parton scatterings.

| $ij \to kl$ | $\lvert\mathcal{M}_{ij\to kl}\rvert^2$ |
|---|---|
| $q_iq_j \to q_iq_j$<br>$q_i\bar{q}_j \to q_i\bar{q}_j$<br>$\bar{q}_iq_j \to \bar{q}_iq_j$<br>$\bar{q}_i\bar{q}_j \to \bar{q}_i\bar{q}_j$ | $\frac{4}{9}g_s^4\frac{s^2+u^2}{t^2},\ i \neq j$ |
| $q_iq_i \to q_iq_i$<br>$\bar{q}_i\bar{q}_i \to \bar{q}_i\bar{q}_i$ | $\frac{4}{9}g_s^4\left(\frac{s^2+u^2}{t^2} + \frac{s^2+t^2}{u^2} - \frac{2}{3}\frac{s^2}{tu}\right)$ |
| $q_i\bar{q}_i \to q_j\bar{q}_j$ | $\frac{4}{9}g_s^4\frac{t^2+u^2}{s^2}$ |
| $q_i\bar{q}_i \to q_i\bar{q}_i$ | $\frac{4}{9}g_s^4\left(\frac{s^2+u^2}{t^2} + \frac{t^2+u^2}{s^2} - \frac{2}{3}\frac{u^2}{st}\right)$ |
| $q\bar{q} \to gg$ | $\frac{8}{3}g_s^4\left(\frac{4}{9}\frac{t^2+u^2}{tu} - \frac{t^2+u^2}{s^2}\right)$ |
| $gq \to gq$<br>$g\bar{q} \to g\bar{q}$ | $g_s^4\left(\frac{s^2+u^2}{t^2} - \frac{4}{9}\frac{s^2+u^2}{su}\right)$ |
| $gg \to q\bar{q}$ | $\frac{3}{8}g_s^4\left(\frac{4}{9}\frac{t^2+u^2}{tu} - \frac{t^2+u^2}{s^2}\right)$ |
| $gg \to gg$ | $\frac{9}{2}g_s^4\left(3 - \frac{tu}{s^2} - \frac{su}{t^2} - \frac{st}{u^2}\right)$ |

Here, $\phi_i$ is the azimuthal angle, $\theta_i$ is the polar angle of a parton's momentum $p_i$, and $\theta_{ij}$ is the polar angle between two partons' momenta $p_i$ and $p_j$, respectively. $f_b$ is the distribution function for thermal partons in the QGP medium. $f_b = \frac{2\cdot 3}{e^{E_2/T}+1}$ follows the Fermi-Dirac distribution for quarks whose spin and color degeneracy is 6, while $f_b = \frac{2\cdot 8}{e^{E_2/T}-1}$ follows the Bose-Einstein distribution for gluons whose degeneracy is 16. The matrix element $|\mathcal{M}_{ab\to cd}|^2$ for two-body scattering is summarized in Table I, which is averaged (summed) over the initial (final) state color and spin [72,81]. $\delta E_{ab\to cd}$ is the energy transfer between the leading parton and the thermal medium parton. Since the masses of high-energy partons are much smaller than their energies, we take the masses of light quarks to be zero, which gives rise to infrared divergences in phase-space integrals. To address these infrared divergences, we adopt a Lorentz-invariant regularization scheme [57],

$$S_2(s,t,u) = \theta\big(s \geqslant 2\mu_D^2\big)\theta\big(-s+\mu_D^2 \leqslant t \leqslant -\mu_D^2\big), \quad (7)$$

where $\mu_D$ is the Debye mass $\mu_D^2 = g^2T^2(N_c/3 + N_f/6)$. $N_f = 3$ is for the light quarks $u$, $d$, and $s$. We employ a non-running fixed strong coupling constant $\alpha_s = g^2/(4\pi)$, which is the only free parameter in our theoretical calculation and influences the strength of both scattering amplitude $|\mathcal{M}|^2$ and Debye mass $\mu_D$ in elastic collision processes, while simultaneously contributing to the radiation spectra of gluons via parton splitting functions and the jet quenching coefficients $\hat{q}$.

In jet reconstruction analyses, the elastic scattering between hard partons and the thermal medium constituents will affect the net energy loss of jet at a given cone size [57]. If the recoil thermal parton falls inside the jet cone after elastic scattering, it will become one part of the jet energy and could decrease the net energy loss for jet. If the recoil thermal parton was inside the jet cone before elastic scattering, one should consider its energy as a part of the initial energy of the jet. When considering the contributions of the recoiling partons before and after elastic scattering, the net energy loss for a jet with jet-cone size $R$ can be expressed as [62]

$$\delta E^{\rm el}_{\rm w/recoil} = E_a + E_b\theta_b - E_c\theta_c - E_d\theta_d, \quad (8)$$

where $\theta_i$ is a $\theta$ function that is related to the relative angle between the partons and jet. If the parton falls inside the jet cone, $\theta_i = 1$, otherwise, $\theta_i = 0$.

### B. Radiative energy loss with $p_{\rm T}$ broadening

As hard partons propagate through the QGP, radiative energy loss due to multiple scattering and induced gluon bremsstrahlung is widely regarded as the dominant mechanism underlying the suppression of single inclusive hadron and jet production [19,31,32,39]. In the study of hadrons, considering the radiated gluons carrying away the energy of the leading partons can well describe the suppression of hadron yields [35–37,41,82]. Most of the radiation energy loss formula are based on the collinear and soft gluon radiation approximations [19,83,84]. For reconstructed jets, the radiation gluons may not be able to escape from the jet cone, thus making no contribution to the net energy loss for jet. Moreover, the further interaction of the radiation gluons also has a probability to change the net energy loss within the jet cone [57,62]. Therefore, in this study for the jet radius dependence of jet suppression, we consider the influence of the transverse momentum broadening effect on the net energy loss of jet with finite cone-size $R$.

We adopt the Higher-Twist (HT) approach to get the radiative energy loss of a leading parton $a$ [17,18,62,63],

$$\frac{\Delta E_a^{\rm rad}}{E} \approx \frac{2C_A\alpha_s}{\pi}\int_{\tau_0}^{\infty} d\tau \int_{\epsilon}^{0.5} dz \int_0^{E^2} \frac{dl_{\rm T}^2}{l_{\rm T}^2(l_{\rm T}^2+\mu_D^2)} \times \hat{q}_a(\tau, \vec{r} + (\tau-\tau_0)\vec{n}) z P_{ga}(z) \sin^2\left(\frac{l_{\rm T}^2(\tau-\tau_0)}{4z(1-z)E}\right), \quad (9)$$

where $C_A = 3$, and $\alpha_s$ is the strong coupling constant as in elastic-scattering processes. $z$ is the longitudinal (along the jet direction) energy fraction and $l_{\rm T}$ is the transverse momentum of the radiated gluon. $P_{ga}(z)$ is the splitting function without the color factor. For a quark $P_{gq}(z) = [1 + (1-z)^2]/z$, and for a gluon $P_{gg} = [1 + z^4 + (1-z)^4]/[z(1-z)]$. The radiative jet energy loss is integrated over the parton propagation path starting from the initial transverse position $\vec{r}$, which is sampled by the Glauber model [85] in Eq. (2), at an initial time $\tau_0 = 0.6$ fm/$c$, and propagating along the direction $\vec{n}$. Since the momentum fraction transferred from the propagating parton to the medium parton during inelastic scattering is much less than 1, we impose the constraint $l_{\rm T}^2 \ll 2E\omega$, which implies a lower bound on $z$, $z \geqslant \epsilon \equiv \frac{1}{2}[1 - \sqrt{1 - 2l_{\rm T}^2/E\omega}]$. Here $E$ is the energy of the hard parton and $\omega$ is the average energy of a medium parton. In addition, we impose the kinematic constraints $l_{\rm T}^2/E^2 \leqslant z^2$ and $l_{\rm T}^2/E^2 \leqslant (1-z)^2$. Furthermore, to restrict to soft-gluon radiation, we limit the $z$ integration to $z \leqslant \frac{1}{2}$ [17,18,63].

$\hat{q}_a$ is the jet transport parameter, which is defined as the average transverse momentum broadening squared per unit length [15]. In recent years, a large number of studies have extracted the values of jet transport coefficients through the model-to-data comparison method [35,36,41–45,47,82,86–89]. In this study, we use a $\hat{q}_a$ form given by leading-order perturbative elastic scattering between parton $a$ and the medium [57,62],

$$\hat{q}_a = C_a \frac{42\zeta(3)}{\pi}\alpha_s^2 T^3 \ln\left(\frac{s^*}{4\mu_D^2}\right), \tag{10}$$

where the color factor $C_a = C_F = 4/3$, $s^* = 5.8ET$ for a quark and $C_a = C_A = 3$, $s^* = 5.6ET$ for a gluon. $\mu_D$ is the Debye mass, and $\zeta(3) \approx 1.202$ is Apéry's constant. The $\hat{q}_a$ in the hydrodynamic QGP fluid, $\hat{q} = \hat{q}(T)\, p^\mu u_\mu / p^0$, depends on both the fluid velocity $u^\mu$ and the temperature $T$ in the local comoving frame, where $p^\mu = (p^0, \vec{p})$ is the four-momentum of the parton. Such a transformation boosts the energy-loss integral back to the collision frame [90]. The dynamical evolution of the QGP medium is provided by the CLVisc (3 + 1)-dimensional hydrodynamic model [91,92]. The initial conditions for the hydrodynamic simulations for each centrality class are obtained by averaging over 200 events from the TRENTo initial-condition model [93]. The overall normalization factors in the TRENTo initial conditions are fixed by comparing with experimental multiplicities at each collision energy. In this work, we assume that jets stop losing energy below the pseudocritical temperature $T_c = 0.165$ GeV.

Due to the jet having a certain cone size and the radiated gluons carrying transverse momentum, we made three modifications to obtain the net energy loss within a given jet cone. The first modification considers the collinear approximation. We assume that gluons radiated collinearly with the jet axis and with radiation angles smaller than the jet radius do not cause net energy loss,

$$l_{\rm T}/zE < \sin R. \tag{11}$$

The second correction concerns soft gluon radiation. We assume that the radiated gluons with energy below the Debye screening mass $\mu_D$ will thermalize into the medium, thus, regardless of their transverse angles, they contribute to a net energy loss for the jet as $zE < \mu_D \sim gT \sim 1$ GeV. The third modification is the transverse momentum broadening effect, which enhances the probability of gluons escaping the jet cone. The average transverse momentum broadening can be obtained by integrating $\hat{q}$ along the jet path,

$$\langle \Delta l_{\rm T}^2 \rangle = \int_{\tau_0}^{\infty} d\tau\, \hat{q}(\vec{r}_0 + \vec{v}\tau). \tag{12}$$

With this additional $\Delta l_{\rm T}$, the final kinetic restriction on the radiated gluons that fall outside jet cone is given by

$$(\vec{l}_{\rm T} + \Delta\vec{l}_{\rm T})/zE < \sin R. \tag{13}$$

We assume that the $\langle \Delta l_{\rm T}^2 \rangle$ broadening distribution caused by radiated gluons follows a Gaussian form [57].

Taking these three modification into Eq. (9), the net energy loss for jet with a finite cone-size $R$ can be obtained by

$$\begin{aligned}\frac{\Delta E_{\rm jet}^{\rm rad}}{E} \approx\ & \frac{2C_A\alpha_s}{\pi}\int_{\tau_0}^{\infty} d\tau \int_{\epsilon}^{0.5} dz \int_0^{E^2} \frac{dl_{\rm T}^2}{l_{\rm T}^2(l_{\rm T}^2+\mu_D^2)} \\ & \times \int d^2\Delta\vec{l}_{\rm T} \frac{1}{2\pi\langle \Delta l_{\rm T}^2\rangle} e^{-\Delta\vec{l}_{\rm T}^{\,2}/(2\langle \Delta l_{\rm T}^2\rangle)} \\ & \times [1-\theta(zE-\mu_D)\theta(zE\sin R - |\vec{l}_{\rm T}+\Delta\vec{l}_{\rm T}|)] \\ & \times \hat{q}_a(\tau, \vec{r}+(\tau-\tau_0)\vec{n})zP_{ga}(z)\sin^2\left(\frac{l_{\rm T}^2(\tau-\tau_0)}{4z(1-z)E}\right).\end{aligned} \tag{14}$$

The $\theta$ function determines the conditions for radiated gluons that cause net energy loss of jets.

### C. Nuclear modification factors

To demonstrate the suppression of the single inclusive jet spectrum in $A+A$ collisions relative to that in $p+p$ collisions, one can define the nuclear modification factor $R_{AA}(p_{\rm T})$ as [94]

$$R_{AA}(p_{\rm T}) = \frac{1}{T_{AA}(\vec{b})}\frac{dN_{AA\to {\rm jet}+X}/dy\, dp_{\rm T}}{d\sigma_{pp\to {\rm jet}+X}/dy\, dp_{\rm T}}, \tag{15}$$

where $T_{AA}(\vec{b}) = \int d^2\vec{r}\, t_A(\vec{r}) t_B(\vec{r}+\vec{b})$ is the overlap function of two colliding nuclei at a given impact parameter $\vec{b}$.

We calculate the $R_{AA}$ for single inclusive jet, and compare the numerical results to the experimental data by utilizing the $\chi^2$/d.o.f. fitting method to determine the free parameter $\alpha_s$. The $\chi^2$/d.o.f. is defined as follows:

$$\chi^2/\text{d.o.f.} = \sum_{i=1}^{N}\left[\frac{(V_{\rm th}-V_{\rm exp})^2}{\sigma_{\rm sys}^2+\sigma_{\rm stat}^2}\right]_i / N, \tag{16}$$

where $V_{\rm th}$ represents the theoretical value, $V_{\rm exp}$ denotes the experimental data, $\sigma_{\rm sys}$ and $\sigma_{\rm stat}$ are the systematic and statistical errors for the experimental data, and $N$ is the number of data points used.

## IV. NUMERICAL RESULTS

Within the framework of the next-to-leading-order (NLO) perturbative QCD (pQCD) parton model, we reconstructed inclusive jets using the anti-$k_{\rm T}$ algorithm [64]. We performed comprehensive calculations and analyses for inclusive jets corresponding to all available experimental data. Building upon the successful establishment of a satisfactory $p+p$ baseline, we calculated the nuclear modification factor ($R_{AA}$) for inclusive jets as a function of transverse momentum ($p_{\rm T}$) and collision centrality. This calculation incorporated the net jet energy loss arising from both collisional processes (elastic scattering with recoiling thermal partons) and radiative energy loss mechanisms (incorporating $p_{\rm T}$ broadening effects) occurring within a given jet cone radius ($R$). Furthermore, we computed the double ratio of $R_{AA}$ values to explicitly illustrate the $R$ dependence of jet suppression induced by these mechanisms.

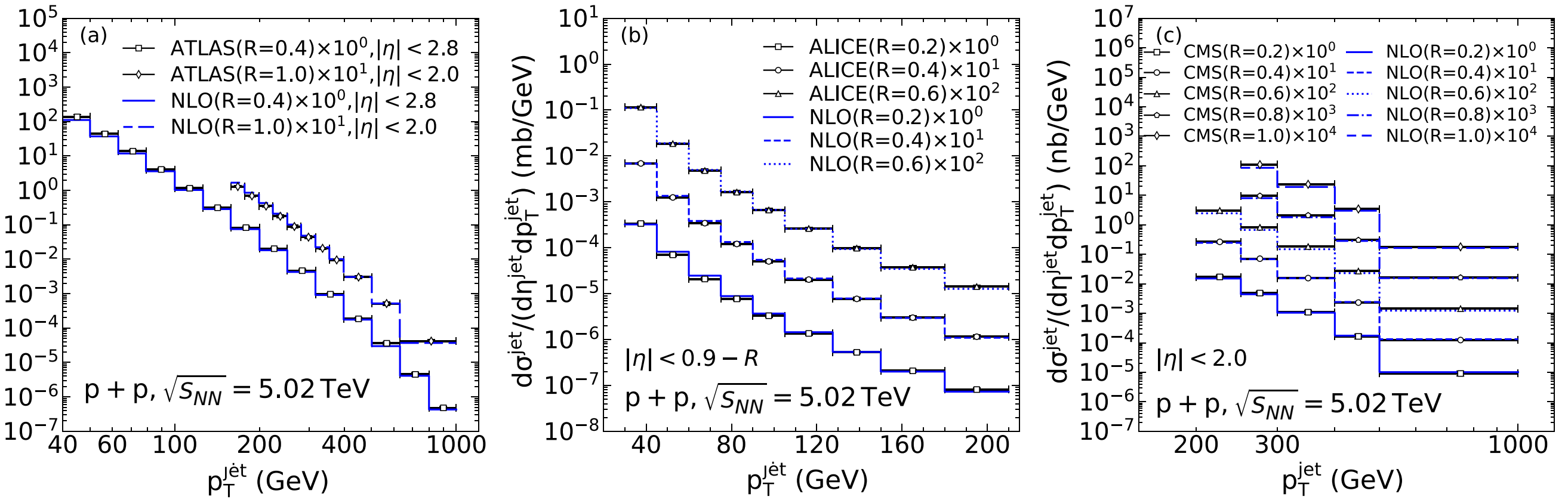

FIG. 1. Single-inclusive jet cross sections for various jet-radius parameters $R$ in $p+p$ collisions at $\sqrt{s_{NN}} = 5.02$ TeV, calculated from NLO pQCD calculations. Jets are reconstructed using the anti-$k_T$ algorithm [64]. For a direct comparison under the same cuts as the experiments, the theory-data comparisons are presented in separate subplots; from left to right corresponding to ATLAS [52,95], ALICE [28], and CMS [30], respectively.

### A. Jet cross-section spectra in $p + p$ collisions

We begin by comparing the single-inclusive jet differential cross sections calculated by NLO pQCD parton model with all available measurements in $p+p$ collisions at $\sqrt{s_{NN}} = 5.02$ TeV, to establish the baseline for jet-quenching studies. Given that the experimental collaborations adopt different measurement criteria, the theory-data comparisons are presented in separate subplots of Fig. 1, corresponding to ATLAS [52,95], ALICE [28], and CMS [30], respectively (left to right). For the ALICE dataset displayed in the middle panel, the measured charged-particle jet transverse-momentum spectra are scaled by a factor of 3/2 to approximate the inclusive-jet yield. This correction accounts for the fraction of jet energy carried by charged particles ($\sim$2/3), with the remainder carried by neutral hadrons ($\sim$1/3). We tune the factorization scale at each jet radius to achieve the best agreement with experimental data. Specifically, for $R =$ 0.2, 0.4, 0.6, 0.8, and 1.0, we take $\mu = 0.4$, 1.0, 2.5, 0.6, and 0.6 $p_{T\max}$, respectively, where $p_{T\max}$ denotes the transverse momentum of the leading parton inside the jet cone. Since the jets are reconstructed at the parton level, without including jet fragmentation or hadronization effects, the optimal scale varies with $R$. Nevertheless, overall the NLO pQCD calculations with the anti-$k_T$ jet reconstruction algorithm demonstrate excellent agreement with the data across a wide range of jet radii ($R = 0.2$–$1.0$) and transverse momenta ($p_T = 40$–$1000$ GeV$/c$).

### B. Inclusive jet suppression

The jet reconstructed in $p+p$ and Pb + Pb collisions is quark or gluon jet. In the NLO pQCD calculations, a jet reconstructed within a specific cone-size may contain two collinear partons: a leading parton and a gluon arising from virtual corrections or real emissions in tree-level diagrams. Thus, the jet flavor becomes ambiguous. We approximate the flavor of the reconstructed jet by the flavor of the leading parton within the jet cone. Given that quarks and gluons lose different amounts of energy in the QCD medium, quantifying the jet energy loss at NLO requires estimating the relative fractions of quark and gluon jets, $f_q$ and $f_g$. The average total energy loss for the reconstructed jet at NLO is then $\Delta E_{\rm jet}^{\rm tot} = (f_q \times \Delta E_{\rm jet}^{q} + f_g \times \Delta E_{\rm jet}^{g})$, where $\Delta E_{\rm jet}^{c} = \Delta E_{\rm jet}^{\rm el} + \Delta E_{\rm jet}^{\rm rad}$ corresponds to Eqs. (8) and (14), and $c$ denotes a quark or a gluon. At LO, a reconstructed jet contains only a single quark or gluon, and we can subtract $\Delta E_{\rm jet}^{c}$ directly from the energy of that parton, i.e., from the jet energy. However, subtracting a flavor-dependent $\Delta E_{\rm jet}^{c}$ or, alternatively, subtracting an average total energy loss $\Delta E_{\rm jet}^{\rm tot}$ (as in the NLO-inspired treatment) should have a similar impact on the suppression of the jet yield. Therefore, in both LO and NLO, we subtract an average total energy loss from the reconstructed jet energy and obtain the jet-yield spectrum in heavy-ion collisions via Eq. (2).

We simulate the quark- and gluon-jet fractions at LO using our pQCD parton model, as shown in Fig. 2. At low jet $p_T$, the gluon fraction is larger. As the jet $p_T$ increases, the fraction of quark rises, reaching about 0.8 at $p_T \sim 1$ TeV. Since gluons lose more energy than quarks in both elastic and radiative channels, the increasing quark-jet fraction with $p_T$ leads to weaker jet suppression at higher $p_T$. Beyond leading order, further splittings and the parton shower evolution of the leading parton can modify the relative fractions of quark and gluon jets, thereby affecting the jet suppression. However, higher-order corrections have a relatively modest impact on these fractions. In Ref. [62], the jet suppression calculated using quark- and gluon-jet fractions from PYTHIA including parton showers [96] is smaller by less than 5% compared to the result obtained with LO pQCD fractions. In addition, the relative fractions of quark and gluon jets predicted by the QCD factorization approach to jet cross sections [97] are also similar to the LO pQCD results. Nevertheless, given the increasing precision of jet-yield suppression measurements, future work should incorporate quark- and gluon-jet fractions that include higher-order corrections.

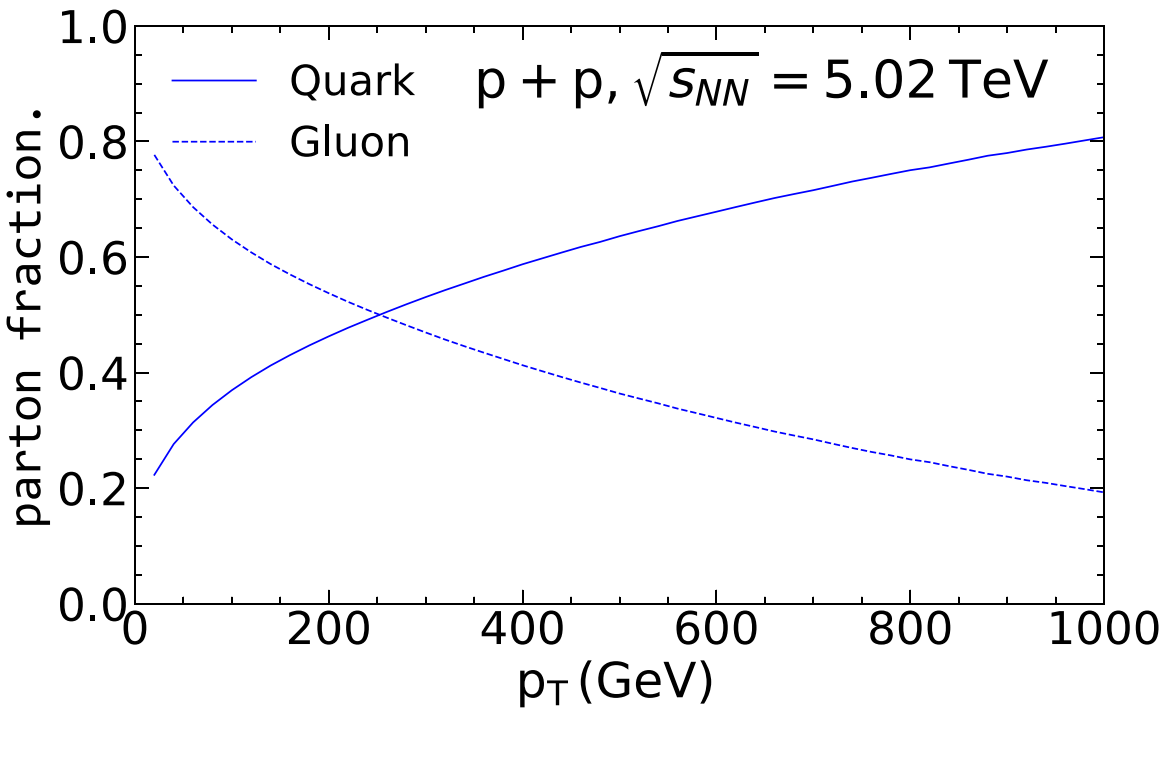

FIG. 2. Quark and gluon fractions as a function of $p_T$ simulated by pQCD parton model at the leading-order approximation in $p+p$ collisions at $\sqrt{s_{NN}} = 5.02$ TeV.

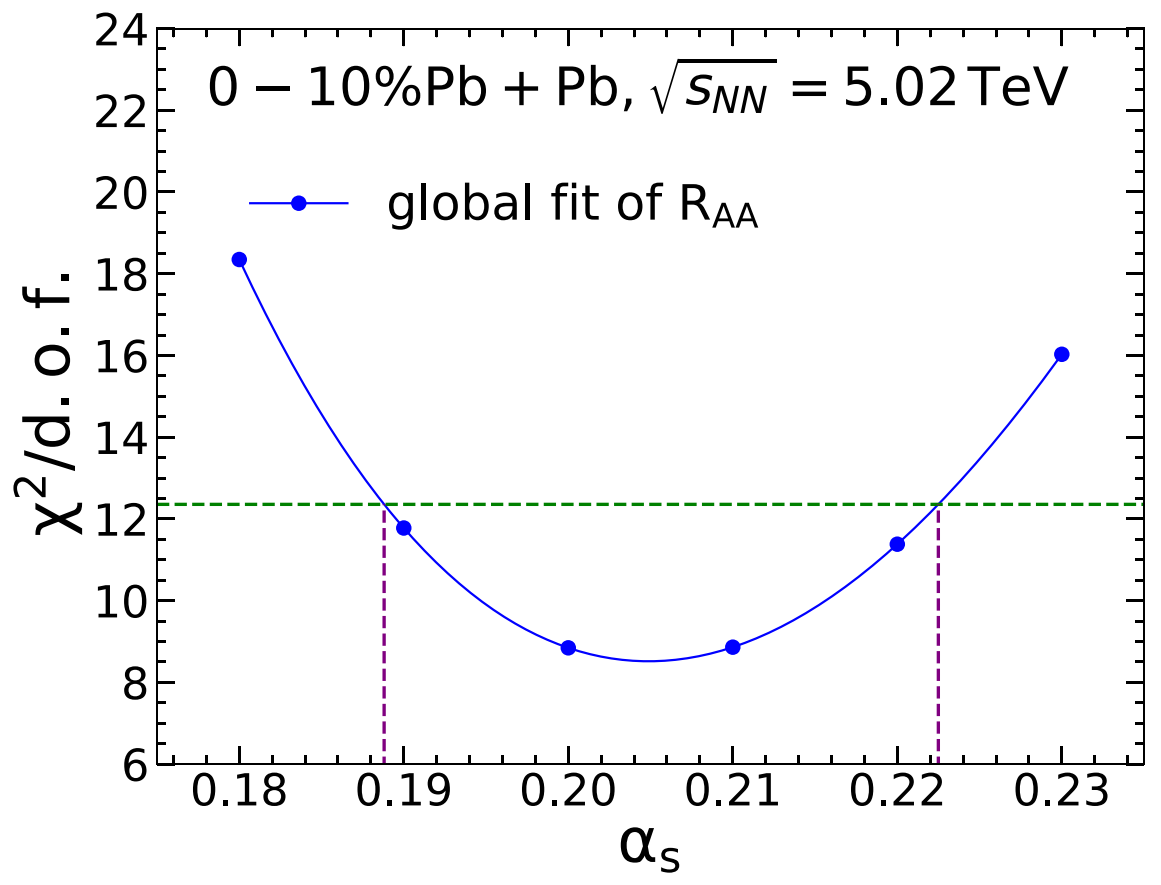

FIG. 3. The global fit $\chi^2$/d.o.f. results from fitting to jet $R_{AA}$ in 0–10% Pb + Pb collisions at $\sqrt{s_{NN}} = 5.02$ TeV.

The jet yield spectrum in $A + A$ collisions is computed by incorporating the fraction-averaged jet energy loss into Eq. (2). We determined the sole free parameter, the strong coupling constant $\alpha_s$, through a global $\chi^2$/d.o.f. comparison against all available experimental data on inclusive jet suppression in Pb + Pb collisions within a given centrality bin. Given that the jet energy loss does not have a simple cubic dependence on the medium temperature [41,43,45], and that the medium temperature spans a broad range for Pb + Pb collisions at 5.02 TeV, we determine the value of $\alpha_s$ separately for each centrality class. Figure 3 presents the resulting $\chi^2$/d.o.f. values obtained from the comparison with jet suppression data in 0–10% central Pb + Pb collisions at $\sqrt{s_{NN}} = 5.02$ TeV. Within the $2\sigma$ confidence interval, we adopt $\alpha_s = 0.189$–$0.223$.

Ideally, one should extract a single, consistent $\alpha_s$ by fitting all jet data across collision centralities and jet radii. However, current studies of the jet transport coefficient $\hat{q}$ indicate that it likely has additional temperature dependence beyond $T^3$ [44,45,47], and different virtuality ranges may require different functional forms in order to describe data from different collision systems, centralities, and momentum intervals simultaneously [98,99]. In this work, we focus on precisely quantifying how different energy-loss mechanisms influence the cone-size dependence of jet suppression. Therefore, we extract best-fit values of $\alpha_s$ independently for different collision centralities. We expect that, after further improving the functional dependence of $\hat{q}$ on $p_{\mathrm{T}}$ and $T$, a single $\alpha_s$ consistent across centralities can be obtained.

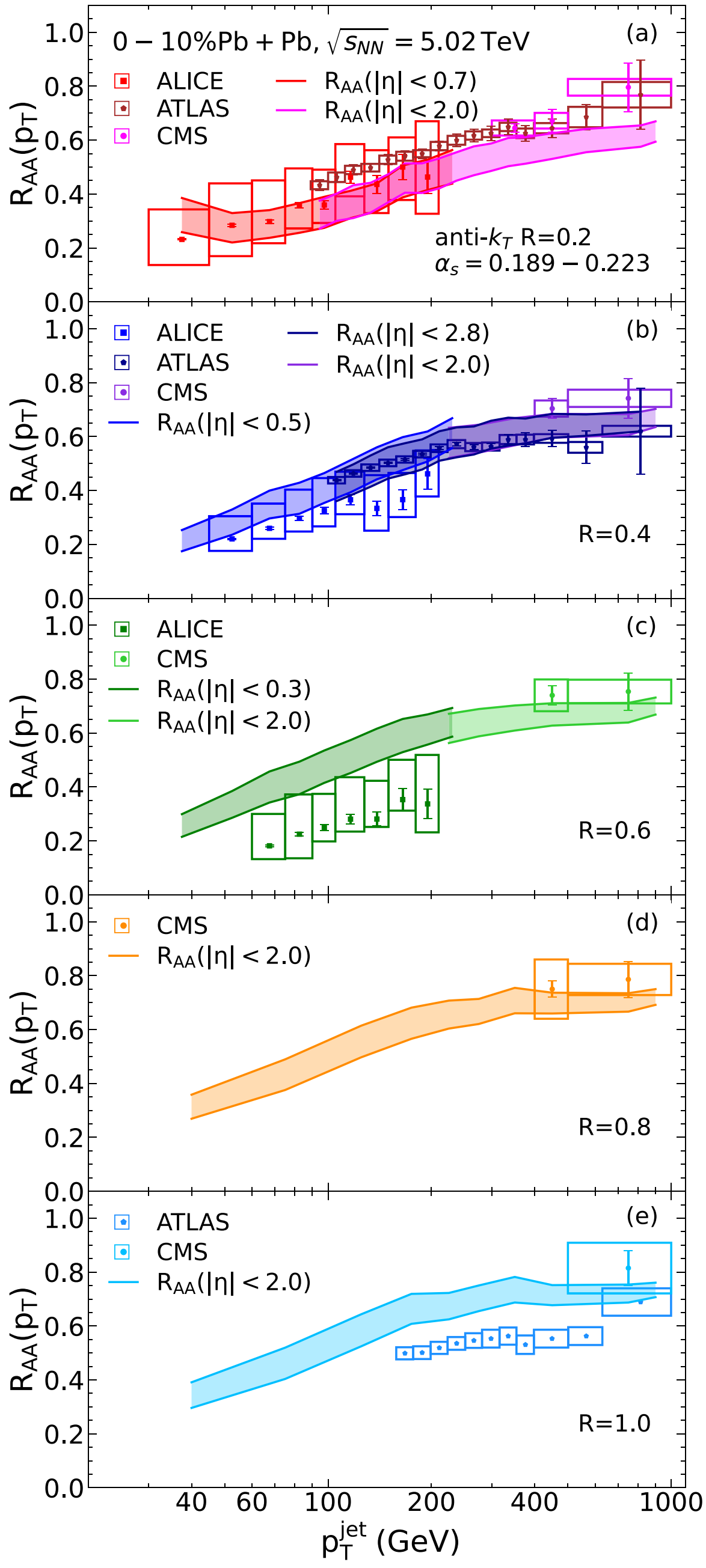

FIG. 4. The nuclear modification factors $R_{AA}$ of single inclusive jet as functions of $p_T$ for various jet radius parameters $R$ in 0–10% Pb + Pb collisions at $\sqrt{s_{NN}} = 5.02$ TeV, compared with measurements from ALICE [28], ATLAS [26], and CMS [30] Collaborations. From top to bottom, the subplots correspond to $R =$ 0.2, 0.4, 0.6, 0.8, and 1.0, respectively.

Using $\alpha_s = 0.189$–$0.223$, we computed the nuclear modification factors $R_{AA}(p_{\mathrm{T}})$ of single inclusive jet for various jet radii $R$ with the NLO pQCD parton model in 0–10% Pb + Pb collisions at $\sqrt{s_{NN}} = 5.02$ TeV. In Fig. 4, from top to bottom, the subplots correspond to $R = 0.2$, 0.4, 0.6, 0.8, and 1.0, respectively. All available experimental data are presented, covering a transverse momentum range from 40 to 1000 GeV [26,28,30]. The numerical results, including collisional energy loss with recoiling thermal partons and radiative energy loss with transverse-momentum broadening effect, agree well with the data within uncertainties, particularly for the smaller radii $R = 0.2$ and 0.4. For larger $R > 0.6$, the jet suppression computed by pQCD model is somewhat weaker than experimental measurements in the intermediate-$p_T$ region, while agreement is restored at high $p_T$. Within our energy loss model, two mechanisms contribute to the reduced suppression observed for large-$R$ jets: as $R$ increases, (i) the probability for radiated soft gluons to escape the jet cone decreases, and (ii) the likelihood that recoiling medium partons fall inside the cone increases. The data, however, exhibit a weaker $R$ dependence than predicted, indicating that the in-cone energy-loss dynamics may require further refinement. Additionally, $R_{AA}$ increases with $p_T$ and approaches unity at very high $p_T$. Our results are consistent with the data in the high-$p_T$ region across all jet radii, reflecting the growing dominance of quark jets at high $p_T$ and the smaller energy loss of quarks relative to gluons.

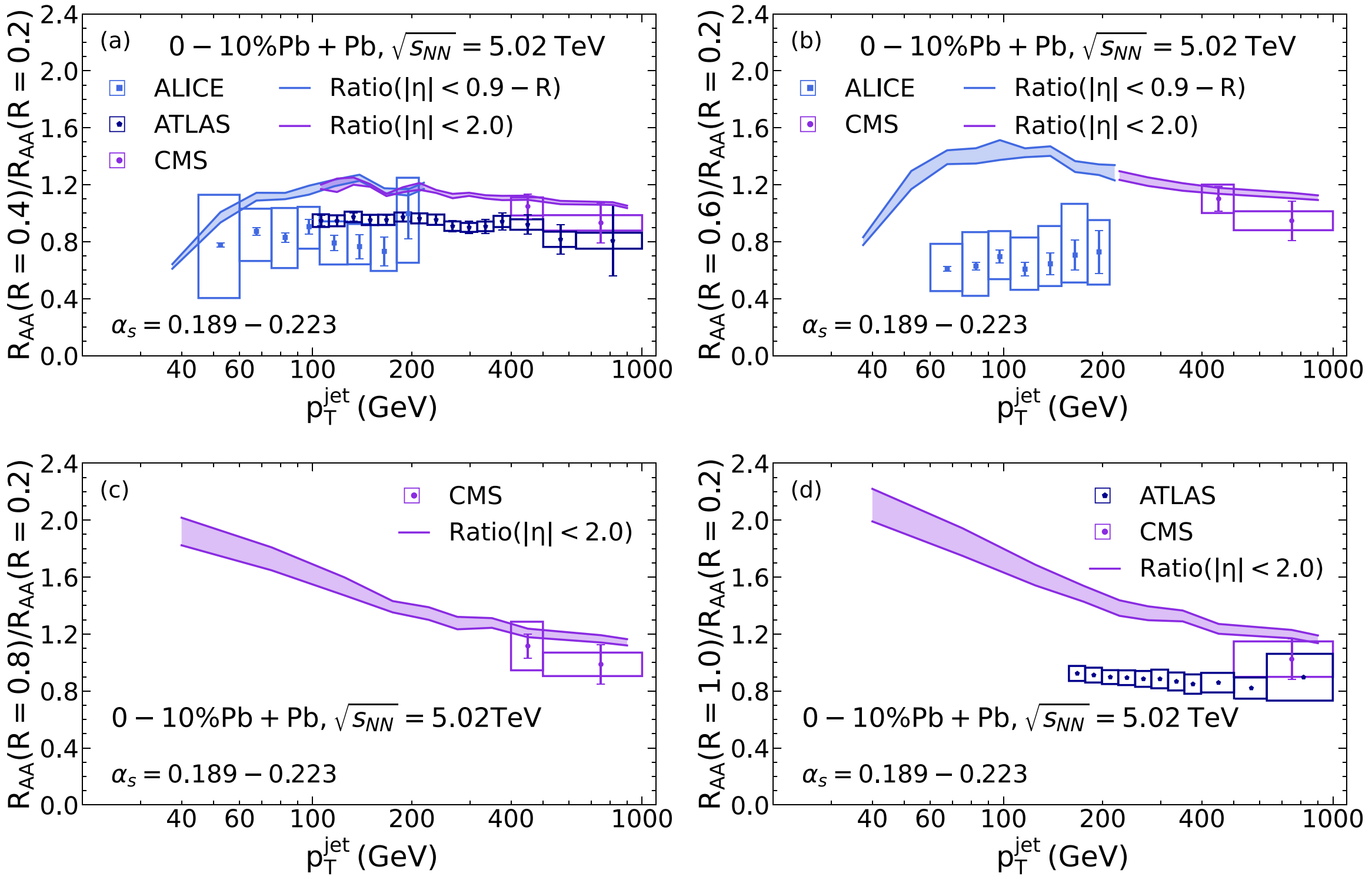

FIG. 5. The double ratios of the single-inclusive jet $R_{AA}$, defined as $R_{AA}(R)/R_{AA}(R=0.2)$ for jet radii $R = 0.4, 0.6, 0.8$, and 1.0, as functions of $p_T$ in 0–10% Pb + Pb collisions at $\sqrt{s_{NN}} = 5.02$ TeV, compared with experimental data [26,28,30].

To more clearly demonstrate how jet suppression depends on the jet radius $R$, we computed the double ratios of $R_{AA}$ with various $R = 0.4$, 0.6, 0.8, and 1.0 to with $R = 0.2$. Figure 5 presents the $p_T$ dependence of these double ratios, compared with experimental data. Uncertainties from varying $\alpha_s$ are largely cancel out and smaller than that shown in Fig. 4. The experimental data for double ratios of $R_{AA}$ are near unity, indicating an $R$-independent jet suppression. Our results are also near to unity for $R_{AA}$ with $R = 0.4$ relative to $R = 0.2$, consistent with the data for small jet radii. However, for $R > 0.6$ relative to $R = 0.2$, the double ratios exceed 1, especially at low and intermediate $p_T < 200$ GeV region. This occurs because a sizable fraction of the lost energy is recovered within the jet cones by capturing the broader angular distribution of radiated gluons and more recoiling thermal partons in this $p_T$ range. At higher $p_T$, jets are less modified by the medium and the additional energy recovered with larger $R$ is reduced. Accordingly, the double ratios approach unity, indicating diminished $R$ sensitivity. The remaining discrepancy with the data suggests that further refinement of in-cone energy-loss and medium-response dynamics is needed to identify the sources of the net energy change within the jet cone.

To further elucidate the $R$ dependence of jet suppression, we compute the net energy loss of a light quark jet with various cone sizes in 0–10% Pb + Pb collisions at 5.02 TeV. The jet is produced at the medium center $x = y = z = 0$ and propagates along $\phi = 0$. As shown in Fig. 6, from top to bottom, the subpanels show the total, radiative and collisional energy loss, respectively. Both collisional and radiative components decrease systematically with increasing jet radius $R$, leading to a monotonic reduction of the total energy loss. In the ultrahigh transverse momentum range, the fraction of energy loss ($\Delta E/E$) becomes small, leading to a convergence of $R_{AA}$ across distinct jet radii $R$. Consequently, the double ratios of $R_{AA}$ tend to unity. As jet radius increases, energy redistributed to larger angles by elastic and radiative processes is more likely to remain inside the jet, so a larger cone recovers more of the energy that would otherwise be lost. In other words, increasing $R$ reduces both the probability that elastically scattered partons escape the jet cone and the chance that radiated gluons fall outside the cone. Consequently, the total energy loss decreases, leading to an enhanced nuclear modification factor $R_{AA}$ and weakened jet suppression. Figure 6 also presents the elastic energy loss calculated without the inclusion of recoil parton contributions, the radiative energy loss calculated without the contribution from transverse momentum broadening, and the total energy loss calculated when both those two effects are neglected. The inclusion of recoil parton contributions leads to a reduction in the elastic energy loss. Conversely, the inclusion of transverse momentum broadening results in an enhancement of the radiative energy loss. When both those two additional contributions are included, the total energy loss exhibits an overall increase.

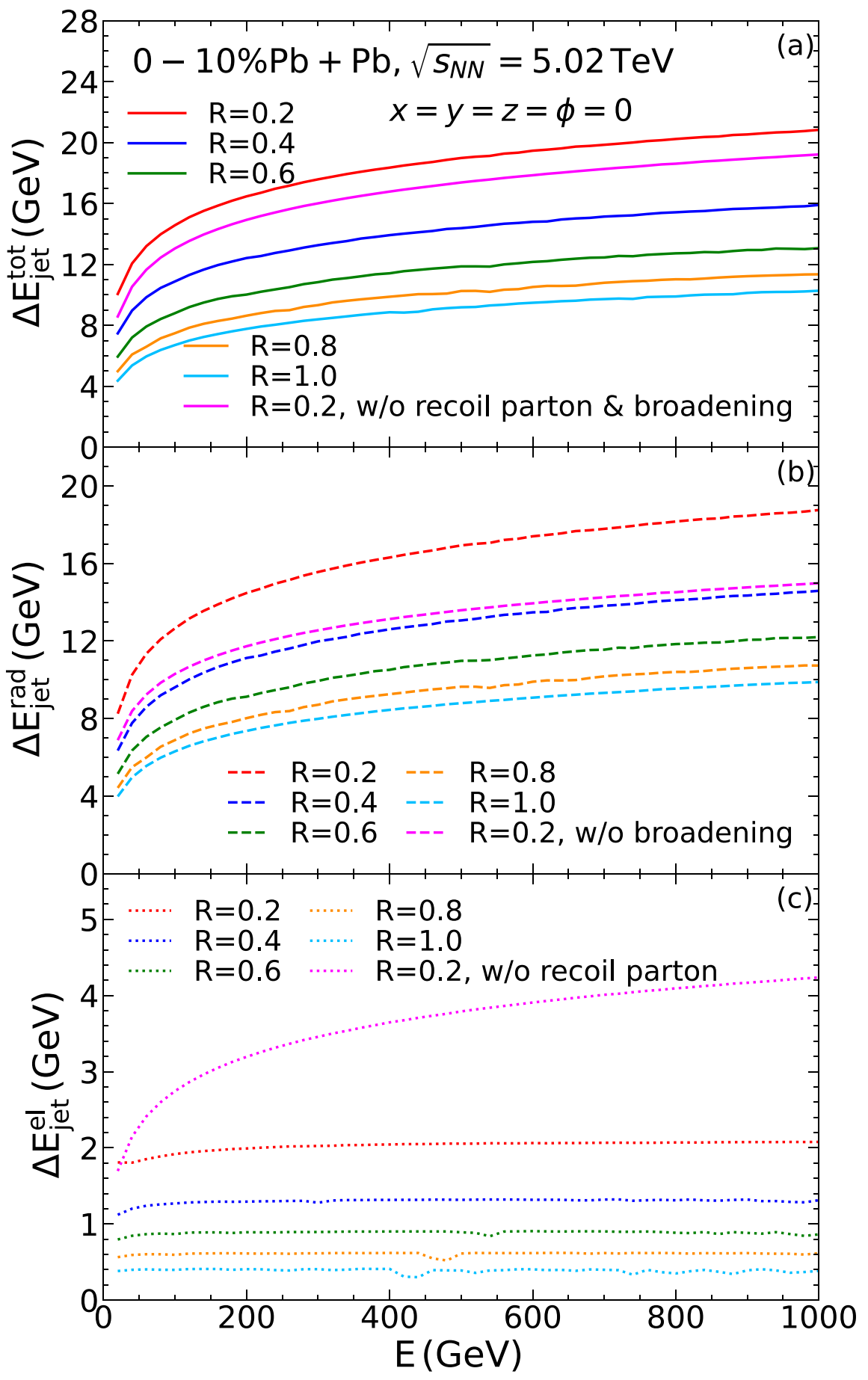

FIG. 6. The energy loss of a light-quark jet with cone size $R = 0.2$–$1.0$ as functions of initial jet energy in 0–10% Pb + Pb collisions at $\sqrt{s_{NN}} = 5.02$ TeV. The jet is produced at the medium center $x = y = z = 0$ and propagates along the $\phi = 0$ direction. From the top to bottom, the subpanels correspond to total, radiative, and collisional energy loss, respectively. For comparison, results without the transverse-momentum broadening in radiative energy loss and without recoil partons in the collisional component are shown as pink curves.

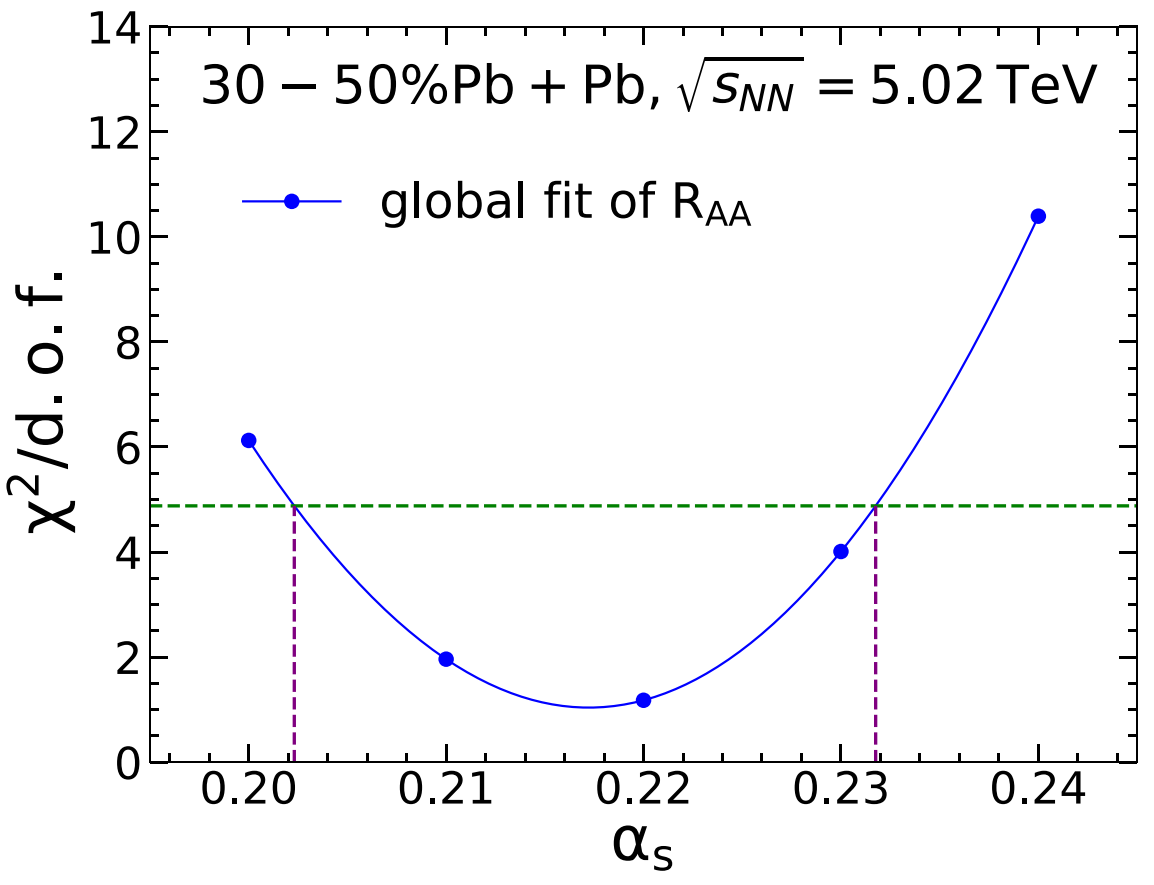

FIG. 7. The global fit $\chi^2$/d.o.f. results from fitting to jet $R_{AA}$ in 30–50% Pb + Pb collisions at $\sqrt{s_{NN}} = 5.02$ TeV.

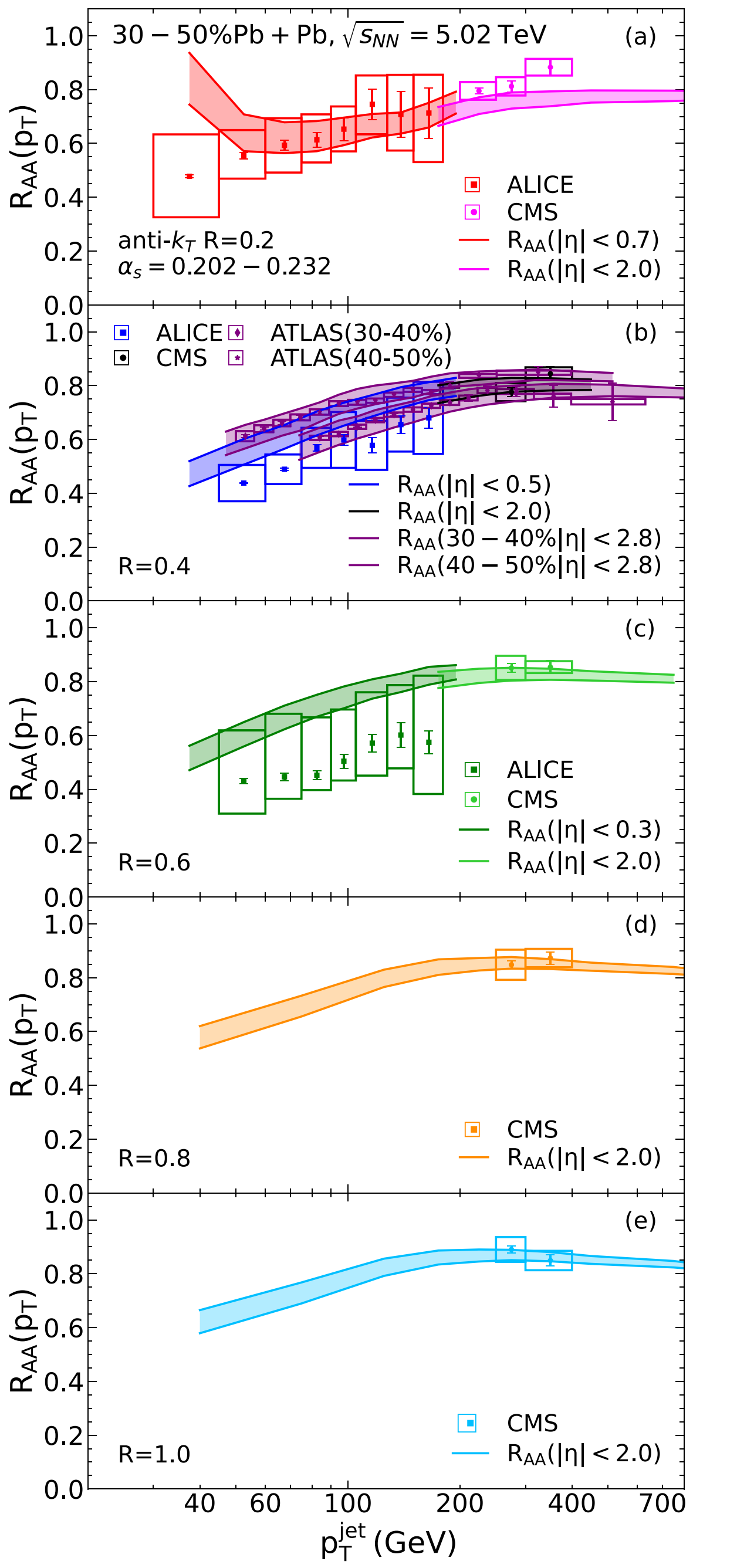

FIG. 8. The nuclear modification factors $R_{AA}$ of single inclusive jet as functions of $p_T$ for various jet radius parameters $R$ in 30–50% Pb + Pb collisions at $\sqrt{s_{NN}} = 5.02$ TeV, compared with measurements from ALICE [28], ATLAS [26], and CMS [30] Collaborations. From top to bottom, the subplots correspond to $R = 0.2$, 0.4, 0.6, 0.8, and 1.0, respectively.

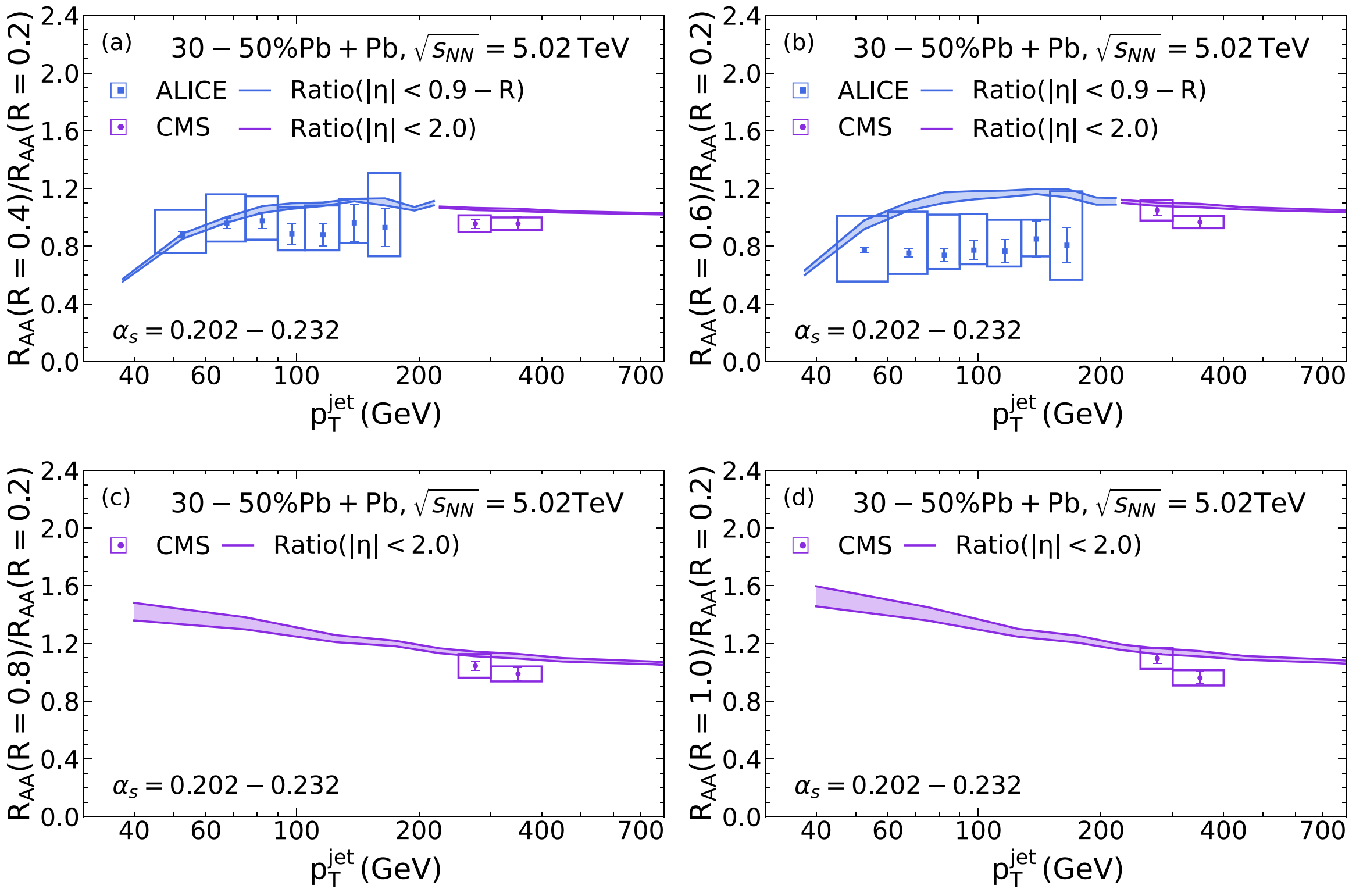

FIG. 9. The double ratios of the single-inclusive jet $R_{AA}$, defined as $R_{AA}(R)/R_{AA}(R=0.2)$ for jet radii $R = 0.4$, 0.6, 0.8, and 1.0, as functions of $p_T$ in 30–50% Pb + Pb collisions at $\sqrt{s_{NN}} = 5.02$ TeV, compared with experimental data [26,28,30].

We further extend the calculations to the 30–50% centrality interval, a region that probes the geometric evolution of the QGP and features a smaller medium size and lower initial energy density and temperature. We calculate the nuclear modification factors $R_{AA}$ of inclusive jet for various radii $R$ in 30–50% Pb + Pb collisions at $\sqrt{s_{NN}} = 5.02$ TeV. The global $\chi^2$/d.o.f. fit to the corresponding experimental data yields a best-fit $\alpha_s$ range of 0.202–0.232, as shown in Fig. 7. The $\alpha_s$ range is slightly larger than that extracted from the 0–10% centrality clas. This observation is consistent with the predicted increase in the jet transport coefficient $\hat{q}$ at lower medium temperatures [41,43,44].With this $\alpha_s$ range, we show the jet $R_{AA}$ in 30–50% Pb + Pb collisions, as depicted in Fig. 8. The resulting $R_{AA}(p_T)$ exhibits weaker suppression for jets than in central collisions, reflecting the reduced energy loss in a smaller, cooler medium. The numerical results are compared to all available experimental data in the 30–50% centrality bin from ALICE [26], ATLAS [26], and CMS [30] groups, covering $p_T = 40$–800 GeV/$c$. Similar to, and in some cases even better than, the results in the 0–10% centrality class, the theoretical results are consistent with the experimental data for all jet radii within the uncertainties, especially in the high-$p_T$ range. This phenomenon further indicates that, in noncentral collisions and at large $p_T$, the sensitivity of the jet suppression on the jet radii is weak.

We also calculate the double ratios $R_{AA}(R)/R_{AA}(R=0.2)$ for $R = 0.4$, $R = 0.6$, $R = 0.8$, and $R = 1.0$ in 30–50% Pb + Pb collisions, as presented in Fig. 9. In these noncentral collisions, the double ratios are near unity across all jet radii at high $p_T$, consistent with the data within uncertainties. At low and intermediate $p_T$, the calculations still show a slight increase of $R_{AA}$ double ratios with increasing $R$. However, experimental coverage for large $R$ in this region is limited. The underlying energy-loss mechanisms are same as for 0–10% centrality.

## V. SUMMARY

In this study, we implement a unified next-to-leading-order pQCD-based parton model that incorporates both elastic and inelastic energy loss mechanisms to investigate the jet modifications in Pb + Pb collisions at $\sqrt{s_{NN}} = 5.02$ TeV. For the collisional energy loss, we include recoil thermal partons, which reduces the net in-cone energy loss for a given jet radius $R$. For the radiative energy loss, computed within a higher-twist formalism, we account for the angular distribution of radiated gluons, the thermalization of soft gluons, and transverse-momentum ($p_T$) broadening effects. Both the collisional and radiative in-cone energy losses decrease with increasing jet radius $R$, leading to a monotonic reduction of the total net jet energy loss.

With these ingredients, we computed the jet nuclear modification factors $R_{AA}$ as a function of $p_T$ for various jet radii $R$. The numerical results are compared with measurements from the ALICE, ATLAS, and CMS Collaborations, covering jet $p_T = 40$–$1000$ GeV/$c$, radii $R = 0.2$–$1.0$, and centrality classes 0–10% and 30–50%. A comprehensive treatment of the energy-loss mechanisms yields weaker jet suppression with increasing jet radius. Moreover, $R_{AA}$ increases with $p_T$ and tends toward unity above $p_T \gtrsim 200$ GeV/$c$. We also evaluate double ratios of jet modification factors, $R_{AA}(R)/R_{AA}(0.2)$, to quantify the $R$ dependence of jet suppression, an observable that is particularly sensitive to changes in the net in-cone energy loss. Numerically, the double ratios are approximately unity for small radius ($R = 0.4$ relative to $R = 0.2$), in agreement with the data within uncertainties. For $R > 0.6$, they exceed unity at low and intermediate $p_T$ and approach unity at high $p_T \gtrsim 200$ GeV/$c$. Increasing $R$ reduces both the probability that elastically scattered partons escape the jet cone and the likelihood that radiated gluons fall outside the jet cone, leading to larger $R_{AA}$ at larger $R$. In addition, the fraction of energy recovered in-cone decreases with increasing jet $p_T$, driving the convergence of $R_{AA}$ across different radii at the high $p_T$. Overall, including recoil partons in the collisional component and transverse-momentum broadening in the radiative component provides a good description of jet suppression for small $R$ and at high $p_T$. For large $R$ ($>0.6$) at intermediate $p_T$ ($<200$ GeV/$c$), further refinement of the microscopic in-cone energy-loss and medium-response mechanisms is still required.

## ACKNOWLEDGMENTS

This work is supported by National Natural Science Foundation of China under Grants No. 12535010 and No. 11935007, and by the Open Fund of Key Laboratory of Quark and Lepton Physics of the Ministry of Education No. QLPL2025P01.

## DATA AVAILABILITY

The data that support the findings of this article are not publicly available upon publication because it is not technically feasible and/or the cost of preparing, depositing, and hosting the data would be prohibitive within the terms of this research project. The data are available from the authors upon reasonable request.

[1] K. Adcox *et al.* (PHENIX Collaboration), Suppression of hadrons with large transverse momentum in central Au + Au collisions at $\sqrt{s_{NN}} = 130$ GeV, Phys. Rev. Lett. **88**, 022301 (2001)..

[2] K. Adcox *et al.* (PHENIX Collaboration), Centrality dependence of the high pT charged hadron suppression in Au + Au collisions at $\sqrt{s_{NN}} = 130$ GeV, Phys. Lett. B **561**, 82 (2003).

[3] S. S. Adler *et al.* (PHENIX Collaboration), Suppressed $\pi^0$ production at large transverse momentum in central Au + Au Collisions at $\sqrt{s_{NN}} = 200$ GeV, Phys. Rev. Lett. **91**, 072301 (2003).

[4] A. Adare *et al.* (PHENIX Collaboration), Neutral pion production with respect to centrality and reaction plane in Au + Au collisions at $\sqrt{s_{NN}} = 200$ GeV, Phys. Rev. C **87**, 034911 (2013).

[5] J. Adams *et al.* (STAR Collaboration), Transverse-momentum and collision-energy dependence of high- $p_T$ hadron suppression in Au + Au collisions at ultrarelativistic energies, Phys. Rev. Lett. **91**, 172302 (2003).

[6] J. Adams *et al.* (STAR Collaboration), Direct observation of dijets in central Au + Au collisions at $\sqrt{s_{NN}} = 200$ GeV, Phys. Rev. Lett. **97**, 162301 (2006).

[7] L. Adamczyk *et al.* (STAR Collaboration), Measurements of jet quenching with semi-inclusive hadron+jet distributions in Au + Au collisions at $\sqrt{s_{NN}} = 200$ GeV, Phys. Rev. C **96**, 024905 (2017).

[8] K. Aamodt *et al.* (ALICE Collaboration), Charged-particle multiplicity density at midrapidity in central Pb-Pb collisions at $\sqrt{s_{NN}} = 2.76$ TeV, Phys. Rev. Lett. **105**, 252301 (2010).

[9] K. Aamodt *et al.* (ALICE Collaboration), Centrality dependence of the charged-particle multiplicity density at midrapidity in Pb-Pb collisions at $\sqrt{s_{NN}} = 2.76$ TeV, Phys. Rev. Lett. **106**, 032301 (2011).

[10] S. Chatrchyan *et al.* (CMS Collaboration), Dependence on pseudorapidity and on centrality of charged hadron production in PbPb collisions at $\sqrt{s_{NN}} = 2.76$ TeV, J. High Energy Phys. 08 (2011) 141.

[11] S. Chatrchyan *et al.* (CMS Collaboration), Centrality dependence of dihadron correlations and azimuthal anisotropy harmonics in PbPb collisions at $\sqrt{s_{NN}} = 2.76$ TeV, Eur. Phys. J. C **72**, 2012 (2012).

[12] G. Aad *et al.* (ATLAS Collaboration), Measurement of the pseudorapidity and transverse momentum dependence of the elliptic flow of charged particles in lead–lead collisions at $\sqrt{s_{NN}} = 2.76$ TeV with the ATLAS detector, Phys. Lett. B **707**, 330 (2012).

[13] G. Aad *et al.* (ATLAS Collaboration), Measurement of the distributions of event-by-event flow harmonics in lead-lead collisions at $\sqrt{s_{NN}} = 2.76$ TeV with the ATLAS detector at the LHC, J. High Energy Phys. 11 (2013) 183.

[14] M. Gyulassy and X.-n. Wang, Multiple collisions and induced gluon bremsstrahlung in QCD, Nucl. Phys. B **420**, 583 (1994).

[15] R. Baier, Y. L. Dokshitzer, A. H. Mueller, S. Peigne, and D. Schiff, Radiative energy loss and $p_\perp$-broadening of high energy partons in nuclei, Nucl. Phys. B **484**, 265 (1997).

[16] M. Gyulassy, P. Levai, and I. Vitev, Non-Abelian energy loss at finite opacity, Phys. Rev. Lett. **85**, 5535 (2000).

[17] X.-f. Guo and X.-N. Wang, Multiple scattering, parton energy loss, and modified fragmentation functions in deeply inelastic $eA$ scattering, Phys. Rev. Lett. **85**, 3591 (2000).

[18] W.-t. Deng and X.-N. Wang, Multiple parton scattering in nuclei: Modified Dokshitzer-Gribov-Lipatov-Altarelli-Parisi (DGLAP) evolution for fragmentation functions, Phys. Rev. C **81**, 024902 (2010).

[19] G.-Y. Qin and X.-N. Wang, Jet quenching in high-energy heavy-ion collisions, Int. J. Mod. Phys. E **24**, 1530014 (2015).

[20] A. Adare *et al.* (PHENIX Collaboration), Suppression pattern of neutral pions at high transverse momentum in Au + Au collisions at $\sqrt{s_{NN}} = 200$ GeV and constraints on medium transport coefficients, Phys. Rev. Lett. **101**, 232301 (2008).

[21] G. Aad *et al.* (ATLAS Collaboration), Measurement of charged-particle spectra in Pb+Pb collisions at $\sqrt{s_{\rm NN}} = 2.76$ TeV with the ATLAS detector at the LHC, J. High Energy Phys. 09 (2015) 050.

[22] V. Khachatryan *et al.* (CMS Collaboration), Charged-particle nuclear modification factors in PbPb and pPb collisions at $\sqrt{s_{NN}} = 5.02$ TeV, J. High Energy Phys. 04 (2017) 039.

[23] S. Acharya *et al.* (ALICE Collaboration), Transverse momentum spectra and nuclear modification factors of charged particles in pp, p-Pb and Pb-Pb collisions at the LHC, J. High Energy Phys. 11 (2018) 013.

[24] J. Adam *et al.* (STAR Collaboration), Measurement of inclusive charged-particle jet production in Au + Au collisions at $\sqrt{s_{NN}} = 200$ GeV, Phys. Rev. C **102**, 054913 (2020).

[25] G. Aad *et al.* (ATLAS Collaboration), Measurements of the nuclear modification factor for jets in Pb + Pb collisions at $\sqrt{s_{NN}} = 2.76$ TeV with the ATLAS detector, Phys. Rev. Lett. **114**, 072302 (2015).

[26] G. Aad *et al.* (ATLAS Collaboration), Measurement of suppression of large-radius jets and its dependence on substructure in Pb + Pb collisions at $\sqrt{s_{NN}} = 5.02$ TeV with the ATLAS Detector, Phys. Rev. Lett. **131**, 172301 (2023).

[27] J. Adam *et al.* (ALICE Collaboration), Measurement of jet suppression in central Pb–Pb collisions at $\sqrt{s_{NN}} = 2.76$ TeV, Phys. Lett. B **746**, 1 (2015).

[28] S. Acharya *et al.* (ALICE Collaboration), Measurement of the radius dependence of charged-particle jet suppression in Pb–Pb collisions at $\sqrt{s_{NN}} = 5.02$ TeV, Phys. Lett. B **849**, 138412 (2024).

[29] V. Khachatryan *et al.* (CMS Collaboration), Measurement of inclusive jet cross sections in *pp* and PbPb collisions at $\sqrt{s_{NN}} = 2.76$ TeV, Phys. Rev. C **96**, 015202 (2017).

[30] A. M. Sirunyan *et al.* (CMS Collaboration), First measurement of large area jet transverse momentum spectra in heavy-ion collisions, J. High Energy Phys. 05 (2021) 284.

[31] M. Gyulassy and M. Plumer, Jet quenching in dense matter, Phys. Lett. B **243**, 432 (1990).

[32] X.-N. Wang and M. Gyulassy, Gluon shadowing and jet quenching in $A + A$ collisions at $\sqrt{s} = 200A$ GeV, Phys. Rev. Lett. **68**, 1480 (1992).

[33] H. Zhang, J. F. Owens, E. Wang, and X.-N. Wang, Tomography of high-energy nuclear collisions with photon-hadron correlations, Phys. Rev. Lett. **103**, 032302 (2009).

[34] H. Zhang, J. F. Owens, E. Wang, and X.-N. Wang, Dihadron tomography of high-energy nuclear collisions in next-to-leading order perturbative QCD, Phys. Rev. Lett. **98**, 212301 (2007).

[35] X.-F. Chen, T. Hirano, E. Wang, X.-N. Wang, and H. Zhang, Suppression of high-$p_T$ hadrons in Pb+Pb collisions at energies available at the CERN Large Hadron Collider, Phys. Rev. C **84**, 034902 (2011).

[36] M. Xie, S.-Y. Wei, G.-Y. Qin, and H.-Z. Zhang, Extracting jet transport coefficient via single hadron and dihadron productions in high-energy heavy-ion collisions, Eur. Phys. J. C **79**, 589 (2019).

[37] M. Xie, X.-N. Wang, and H.-Z. Zhang, $\gamma$ -hadron spectra in $p +$ Pb collisions at $\sqrt{s_{NN}} = 5.02$ TeV, Phys. Rev. C **103**, 034911 (2021).

[38] B. Schenke, C. Gale, and S. Jeon, MARTINI: An event generator for relativistic heavy-ion collisions, Phys. Rev. C **80**, 054913 (2009).

[39] G.-Y. Qin, J. Ruppert, C. Gale, S. Jeon, G. D. Moore, and M. G. Mustafa, Radiative and collisional jet energy loss in the quark-gluon plasma at the BNL relativistic heavy ion collider, Phys. Rev. Lett. **100**, 072301 (2008).

[40] J. Xu, J. Liao, and M. Gyulassy, Consistency of perfect fluidity and jet quenching in semi-quark-gluon monopole plasmas, Chin. Phys. Lett. **32**, 092501 (2015).

[41] K. M. Burke *et al.* (JET Collaboration), Extracting the jet transport coefficient from jet quenching in high-energy heavy-ion collisions, Phys. Rev. C **90**, 014909 (2014).

[42] S. Shi, J. Liao, and M. Gyulassy, Global constraints from RHIC and LHC on transport properties of QCD fluids in CUJET/CIBJET framework, Chin. Phys. C **43**, 044101 (2019).

[43] A. Kumar *et al.* (JETSCAPE Collaboration), Inclusive jet and hadron suppression in a multistage approach, Phys. Rev. C **107**, 034911 (2023).

[44] M. Xie, W. Ke, H. Zhang, and X.-N. Wang, Global constraint on the jet transport coefficient from single-hadron, dihadron, and $\gamma$-hadron spectra in high-energy heavy-ion collisions, Phys. Rev. C **109**, 064917 (2024).

[45] M. Xie, W. Ke, H. Zhang, and X.-N. Wang, Information-field-based global Bayesian inference of the jet transport coefficient, Phys. Rev. C **108**, L011901 (2023).

[46] Y. He, S. Cao, W. Chen, T. Luo, L.-G. Pang, and X.-N. Wang, Interplaying mechanisms behind single inclusive jet suppression in heavy-ion collisions, Phys. Rev. C **99**, 054911 (2019).

[47] M. Xie, Q.-F. Han, E.-K. Wang, B.-W. Zhang, and H.-Z. Zhang, The medium-temperature dependence of jet transport coefficient in high-energy nucleus–nucleus collisions, Nucl. Sci. Tech. **35**, 125 (2024).

[48] Q.-F. Han, M. Xie, and H.-Z. Zhang, Extracting the jet transport coefficient from hadron suppressions by confronting current NLO parton fragmentation functions, Eur. Phys. J. Plus **137**, 1056 (2022).

[49] G. Aad *et al.* (ATLAS Collaboration), Measurement of the jet radius and transverse momentum dependence of inclusive jet suppression in lead–lead collisions at $\sqrt{s_{NN}} = 2.76$ TeV with the ATLAS detector, Phys. Lett. B **719**, 220 (2013).

[50] B. Abelev *et al.* (ALICE Collaboration), Measurement of charged jet suppression in Pb-Pb collisions at $\sqrt{s_{NN}} = 2.76$ TeV, J. High Energy Phys. 03 (2014) 013.

[51] S. Acharya *et al.* (ALICE Collaboration), Measurements of inclusive jet spectra in *pp* and central Pb-Pb collisions at $\sqrt{s_{NN}} = 5.02$ TeV, Phys. Rev. C **101**, 034911 (2020).

[52] M. Aaboud *et al.* (ATLAS Collaboration), Measurement of the nuclear modification factor for inclusive jets in Pb+Pb collisions at $\sqrt{s_{NN}} = 5.02$ TeV with the ATLAS detector, Phys. Lett. B **790**, 108 (2019).

[53] Y.-T. Chien and I. Vitev, Towards the understanding of jet shapes and cross sections in heavy ion collisions using soft-collinear effective theory, J. High Energy Phys. 05 (2016) 023.

[54] Z.-B. Kang, F. Ringer, and I. Vitev, Inclusive production of small radius jets in heavy-ion collisions, Phys. Lett. B **769**, 242 (2017).

[55] K. C. Zapp, F. Krauss, and U. A. Wiedemann, A perturbative framework for jet quenching, J. High Energy Phys. 03 (2013) 080.
[56] R. Kunnawalkam Elayavalli and K. C. Zapp, Medium response in JEWEL and its impact on jet shape observables in heavy ion collisions, J. High Energy Phys. 07 (2017) 141.
[57] Y. He, T. Luo, X.-N. Wang, and Y. Zhu, Linear Boltzmann transport for jet propagation in the quark-gluon plasma: Elastic processes and medium recoil, Phys. Rev. C **91**, 054908 (2015); **97**, 019902(E) (2018).
[58] D. Pablos, Jet suppression from a small to intermediate to large radius, Phys. Rev. Lett. **124**, 052301 (2020).
[59] Y. Mehtar-Tani, D. Pablos, and K. Tywoniuk, Cone-size dependence of jet suppression in heavy-ion collisions, Phys. Rev. Lett. **127**, 252301 (2021).
[60] A. Takacs and K. Tywoniuk, Quenching effects in the cumulative jet spectrum, J. High Energy Phys. 10 (2021) 038.
[61] Y. Mehtar-Tani, D. Pablos, and K. Tywoniuk, Jet suppression and azimuthal anisotropy from RHIC to LHC, Phys. Rev. D **110**, 014009 (2024).
[62] X.-N. Wang, S.-Y. Wei, and H.-Z. Zhang, Effect of medium recoil and $p_T$ broadening on single inclusive jet suppression in high-energy heavy-ion collisions in the high-twist approach, Phys. Rev. C **96**, 034903 (2017).
[63] X.-N. Wang and X.-f. Guo, Multiple parton scattering in nuclei: Parton energy loss, Nucl. Phys. A **696**, 788 (2001).
[64] M. Cacciari, G. P. Salam, and G. Soyez, The anti-$k_t$ jet clustering algorithm, J. High Energy Phys. 04 (2008) 063.
[65] N. Kidonakis and J. F. Owens, Effects of higher-order threshold corrections in high-$E_T$ jet production, Phys. Rev. D **63**, 054019 (2001).
[66] D. de Florian, P. Hinderer, A. Mukherjee, F. Ringer, and W. Vogelsang, Approximate next-to-next-to-leading order corrections to hadronic jet production, Phys. Rev. Lett. **112**, 082001 (2014).
[67] A. Banfi, G. P. Salam, and G. Zanderighi, Infrared-safe definition of jet flavour, Eur. Phys. J. C **47**, 113 (2006).
[68] Z.-B. Kang, F. Ringer, and I. Vitev, The semi-inclusive jet function in SCET and small radius resummation for inclusive jet production, J. High Energy Phys. 10 (2016) 125.
[69] M. Dasgupta, F. Dreyer, G. P. Salam, and G. Soyez, Small-radius jets to all orders in QCD, J. High Energy Phys. 04 (2015) 039.
[70] M. Dasgupta, F. A. Dreyer, G. P. Salam, and G. Soyez, Inclusive jet spectrum for small-radius jets, J. High Energy Phys. 06 (2016) 057.
[71] J. Casalderrey-Solana, D. Gulhan, G. Milhano, D. Pablos, and K. Rajagopal, Angular structure of jet quenching within a hybrid strong/weak coupling model, J. High Energy Phys. 03 (2017) 135.
[72] J. F. Owens, Large-momentum-transfer production of direct photons, jets, and particles, Rev. Mod. Phys. **59**, 465 (1987).
[73] T.-J. Hou *et al.*, New CTEQ global analysis of quantum chromodynamics with high-precision data from the LHC, Phys. Rev. D **103**, 014013 (2021).
[74] B. W. Harris and J. F. Owens, Two cutoff phase space slicing method, Phys. Rev. D **65**, 094032 (2002).
[75] P. Jacobs and G. Cooper, Spatial distribution of initial interactions in high energy collisions of heavy nuclei, arXiv:nucl-ex/0008015.
[76] X.-N. Wang, A pQCD-based approach to parton production and equilibration in high-energy nuclear collisions, Phys. Rep. **280**, 287 (1997).
[77] S.-y. Li and X.-N. Wang, Gluon shadowing and hadron production at RHIC, Phys. Lett. B **527**, 85 (2002).
[78] V. Emel'yanov, A. Khodinov, S. R. Klein, and R. Vogt, Effect of shadowing on initial conditions, transverse energy, and hard probes in ultrarelativistic heavy ion collisions, Phys. Rev. C **61**, 044904 (2000).
[79] T. Hirano and Y. Nara, Interplay between soft and hard hadronic components for identified hadrons in relativistic heavy ion collisions, Phys. Rev. C **69**, 034908 (2004).
[80] K. J. Eskola, P. Paakkinen, H. Paukkunen, and C. A. Salgado, EPPS21: A global QCD analysis of nuclear PDFs, Eur. Phys. J. C **82**, 413 (2022).
[81] E. Eichten, I. Hinchliffe, K. D. Lane, and C. Quigg, Supercollider physics, Rev. Mod. Phys. **56**, 579 (1984); **58**, 1065(E) (1986).
[82] Z.-Q. Liu, H. Zhang, B.-W. Zhang, and E. Wang, Quantifying jet transport properties via large $p_T$ hadron production, Eur. Phys. J. C **76**, 20 (2016).
[83] B.-W. Zhang and X.-N. Wang, Multiple parton scattering in nuclei: Beyond helicity amplitude approximation, Nucl. Phys. A **720**, 429 (2003).
[84] B.-W. Zhang, E.-k. Wang, and X.-N. Wang, Multiple parton scattering in nuclei: Heavy quark energy loss and modified fragmentation functions, Nucl. Phys. A **757**, 493 (2005).
[85] M. L. Miller, K. Reygers, S. J. Sanders, and P. Steinberg, Glauber modeling in high-energy nuclear collisions, Annu. Rev. Nucl. Part. Sci. **57**, 205 (2007).
[86] S. K. Das, F. Scardina, S. Plumari, and V. Greco, Toward a solution to the $R_{AA}$ and $v_2$ puzzle for heavy quarks, Phys. Lett. B **747**, 260 (2015).
[87] S. Cao, L.-G. Pang, T. Luo, Y. He, G.-Y. Qin, and X.-N. Wang, $R_{AA}$ vs. $v_2$ of heavy and light hadrons within a linear Boltzmann transport model, Nucl. Part. Phys. Proc. **289-290**, 217 (2017).
[88] J. Xu, J. Liao, and M. Gyulassy, Bridging soft-hard transport properties of quark-gluon plasmas with CUJET3.0, J. High Energy Phys. 02 (2016) 169.
[89] W. Ke and X.-N. Wang, QGP modification to single inclusive jets in a calibrated transport model, J. High Energy Phys. 05 (2021) 041.
[90] L.-G. Pang, H. Petersen, and X.-N. Wang, Pseudorapidity distribution and decorrelation of anisotropic flow within the open-computing-language implementation CLVisc hydrodynamics, Phys. Rev. C **97**, 064918 (2018).
[91] L. Pang, Q. Wang, and X.-N. Wang, Effects of initial flow velocity fluctuation in event-by-event (3+1)D hydrodynamics, Phys. Rev. C **86**, 024911 (2012).
[92] L.-G. Pang, Y. Hatta, X.-N. Wang, and B.-W. Xiao, Analytical and numerical Gubser solutions of the second-order hydrodynamics, Phys. Rev. D **91**, 074027 (2015).
[93] J. S. Moreland, J. E. Bernhard, and S. A. Bass, Alternative ansatz to wounded nucleon and binary collision scaling in high-energy nuclear collisions, Phys. Rev. C **92**, 011901 (2015).
[94] X.-N. Wang, Energy dependence of jet quenching and lifetime of dense matter in high-energy heavy-ion collisions, Phys. Rev. C **70**, 031901 (2004).
[95] ATLAS Collaboration, Measurement of suppression of large-radius jets and its dependence on substructure in Pb+Pb at 5.02

TeV by ATLAS detector, CERN Report No. ATLAS-CONF-2019-056, 2019 (unpublished).

[96] T. Sjöstrand, S. Mrenna, and P. Z. Skands, A brief introduction to PYTHIA 8.1, Comput. Phys. Commun. **178**, 852 (2008).

[97] J.-W. Qiu, F. Ringer, N. Sato, and P. Zurita, Factorization of jet cross sections in heavy-ion collisions, Phys. Rev. Lett. **122**, 252301 (2019).

[98] S. Cao *et al.* (JETSCAPE Collaboration), Determining the jet transport coefficient $\hat{q}$ from inclusive hadron suppression measurements using Bayesian parameter estimation, Phys. Rev. C **104**, 024905 (2021).

[99] R. Ehlers *et al.* (JETSCAPE Collaboration), Bayesian inference analysis of jet quenching using inclusive jet and hadron suppression measurements, Phys. Rev. C **111**, 054913 (2025).